\documentclass[12pt]{iopart}

\usepackage{graphicx}
\usepackage{color}
\usepackage{amssymb,amssymb}

\begin{document}

\title[Uncertainty relations and topological-band insulator transitions]{Entropic uncertainty  relations and topological-band insulator transitions in 2D gapped Dirac materials}

\author{E. Romera}
\address{Departamento de F\'{\i}sica At\'omica, Molecular y Nuclear and
Instituto Carlos I de F{\'\i}sica Te\'orica y
Computacional, Universidad de Granada, Fuentenueva s/n, 18071 Granada,
Spain}

\author{M. Calixto}
\address{Departamento de Matem\'atica Aplicada, Universidad de Granada,
Fuentenueva s/n, 18071 Granada, Spain}

\begin{abstract}
Uncertainty relations are studied  for a characterization of topological-band insulator transitions in  2D gapped Dirac materials isoestructural
 with graphene.
We show that the relative or Kullback-Leibler entropy in position and momentum spaces,  and the standard variance-based uncertainty relation give sharp 
signatures of topological phase transitions  in these systems.
\end{abstract}

\maketitle

\section{Introduction}

Recently, there is a growing interest in the study of 2D gapped  Dirac materials isoestructural with graphene. 
One of these materials is silicene, which
 is a two dimensional crystal of silicon, with a relevant intrinsic spin-orbit coupling (as compared to graphene), 
studied theoretically \cite{takeda,guzman94} and  experimentally
 \cite{vogt12,Aufray10,lalmi10,fleurence12,padova}. Other gapped  Dirac materials are germanene, stannene and Pb \cite{nature}.
For these systems, the low energy
electronic properties can be described by a Dirac Hamiltonian, like in graphene, but the
electrons are massive  due to the relative
large spin-orbit coupling $\Delta_\mathrm{so}$. In fact, we will consider the application of a  perpendicular electric field
 $\mathcal{E}_z=\Delta_z/l$ ($l$ is the inter-lattice distance
 of the buckled honeycomb structure) to the material sheet, which generates 
 a tunable band gap $|\Delta_{s\xi}|=|(\Delta_z-s\xi\Delta_{\mathrm{so}})/2|$
 ($s$ and $\xi$ denote spin and valley, respectively).
There is  a topological phase transition  \cite{tahir2013}  from a topological  
insulator (TI, $|\Delta_z|<\Delta_\mathrm{so}$) to a band insulator (BI, $|\Delta_z|>\Delta_\mathrm{so}$), at a charge neutrality
point (CNP)  $\Delta_z^{(0)}=s\xi\Delta_\mathrm{so}$, where there is a gap cancellation  between the perpendicular electric field
and the spin-orbit coupling, thus exhibiting a semimetal behavior.

 A 2D topological insulator  was theoretically studied in \cite{Kane} and first 
discovered experimentally in HgTe quantum wells in \cite{Bernevig}. A TI-BI transition is characterized by a band inversion with
 a level crossing at some 
critical value of a control parameter (electric field, quantum well thickness,
etc). Recently,  we have found that electron-hole Wehrl entropies 
of a quantum state in a coherent-state representation 
provide a useful tool to identify  TI-BI  phase transitions \cite{calixto14}.

In this work we  explore the connection between the TI-BI transitions in some 2D Dirac materials
with the Shannon information entropies of the 
wave packet probability densities  in position and momentum spaces and, in particular, with the entropic uncertainty relation. 
The uncertainty principle
can be quantified  in terms of the usual variance-based uncertainty relation, $\Delta x \Delta p\geq\frac12$, or alternatively 
by means of the entropic uncertainty relation \cite{10,11,21}  which  
has been shown to be 
more appropriate  in different physical situations \cite{15,16,Majernik,17,18,found,gadre,sears,calixto12}. 
If we define the position and momentum densities of a state $\Psi$ as $\rho({\bf r})= |\Psi({\bf r})|^2$ and
 $\gamma({\bf p})=|\Phi({\bf p})|^2$, respectively, with  $\Psi({\bf r})$  the position and
 $\Phi({\bf p})$ the momentum wave packets,
the entropic uncertainty relation is given by 
\begin{equation}
 S_{\rho} + S_{\gamma}\geq D(1+\ln\pi) \label{entropic}
\end{equation}
where $D$ is the dimension of the position and momentum space and where 
$S_f=-\int f({\bf x})\ln(f({\bf x})) d{\bf x}$ is the so called Shannon information entropy of 
a density $f$. The equality is reached when the wave packets in position and momentum spaces are Gaussians. 
The Shannon information entropy  measures the uncertainty in the localization of the wave packet
in position or momentum spaces, so that the higher the Shannon entropy is, the smaller  the localization of the wave packet
is; and the smaller the entropy is, the more concentrated the wave function is. Besides we will consider 
the relative or Kullback-Leibler entropy to characterize the topological-band insulator transitions (see section \ref{KLsection}) in these materials.

 The paper is organized as
follows. Firstly, in Section \ref{Hamilsec}, we shall introduce the low energy Hamiltonian for
some 2D Dirac materials (namely: silicene, germanene, stantene,...). Then, in Section
\ref{entropysec}, we will characterize topological-band insulator transitions
in silicene in terms of the entropic uncertainty relation. In Section \ref{KLsection},  the connection between the relative entropy and 
the topological-band insulator transitions is studied.  In Section \ref{numsec} we use the Heisenberg uncertainty relation to characterize the TI-BI phase transition. 
Finally, some concluding
remarks will be given in the last Section.

\section{Low energy Hamiltonian}\label{Hamilsec}

Let us consider a monolayer silicene film with  external magnetic   $B$ and electric
 $\mathcal{E}_z$ fields  applied perpendicular to the silicene plane. The low energy effective Hamiltonian in the vicinity of the Dirac point is given by \cite{tahir2013}
\begin{equation}
H_s^{\xi}=v_F (\sigma_x  p_x-\xi \sigma_y  p_y )-\xi
s \Delta_{\mathrm{so}} \sigma_z+ \Delta_z \sigma_z,
\label{hamiltoniano}
\end{equation}
where $\xi$ corresponds  to the inequivalent corners ${K}$ ($\xi=1$) and
${K}^{\prime}$  ($\xi=-1$) of the Brillouin zone, respectively, 
${\sigma}_j$ are the usual Pauli matrices, $v_F$ is the Fermi
velocity of the Dirac fermions  (see  Table \ref{tabla} for theoretical estimations in Si, Ge and Sn), spin up and down values are 
represented by $s=\pm 1$, respectively, and  $\Delta_{\rm so}$ is the band gap  induced by
intrinsic spin-orbit interaction, which provides a mass to the Dirac fermions. 
We are considering the application of a  constant electric field $\mathcal{E}_z$ which creates a potential difference $\Delta_z=l\mathcal{E}_z$ 
between sub-lattices. The value $l$ appears in table 
\ref{tabla} for different materials.
 The values of the spin-orbit energy gap induced by the intrinsic spin-orbit coupling has been theoretically estimated 
\cite{drummond2012,liu2011,liub2011,nature}
for different 2D Dirac materials that we show in Table \ref{tabla}.

\begin{table} \begin{center}
 \begin{tabular}{|c|c|c|c|}
  \hline
 & $\Delta_{\mathrm{so}}$ (meV) & $l$ ($\AA$)& $v_F$ ($10^{5}$m/s)\\ 
\hline
Si & 4.2 & 0.22& 4.2 \\
Ge & 11.8 & 0.34& 8.8\\
Sn & 36.0 &0.42 & 9.7 \\
Pb & 207.3 & 0.44 & --\\
\hline
 \end{tabular}
 \end{center}
\caption{\label{tabla} Approximate values of model parameters $\Delta_{\mathrm{so}}$ (spin-orbit coupling), $l$ (interlattice distance) and $v_F$ 
(Fermi velocity) for two dimensional 
Si, Ge, Sn and Pb sheets. These data have been obtain from first-principles computations in \cite{nature} 
($\Delta_{\mathrm{so}}$ and $l$) and \cite{fermispeed1,fermispeed2} ($v_F$).}
\end{table}

The eigenvalue problem can be easily solved. Using the Landau gauge, $\vec{A}=(0, Bx,0)$, the corresponding eigenvalues 
and eigenvectors for the $K$  and $K'$ points   are given by
 \cite{tahir2013,stille2012,calixto14}
\begin{equation}
E_{n}^{s\xi}=\left\{\begin{array}{l} \mathrm{sgn}(n) \sqrt{|n|\hbar^2\omega^2 + \Delta_{s\xi}^2}, \quad n\neq 0, \\ 
              -\xi\Delta_{s\xi}, \quad n= 0, \end{array}\right.\label{especteq}
            \end{equation}
 and
\begin{equation}
|n\rangle_{s\xi}=\left(\begin{array}{c}
-i A_{n}^{s\xi}||n|-\xi_+\rangle\\
B_{n}^{s\xi}||n|-\xi_-\rangle
\end{array}\right),
\label{vectors}
\end{equation}
where we denote by $\xi_\pm=(1\pm\xi)/2$, the Landau level index $n=0,\pm 1,\pm 2,\dots$, the cyclotron frequency 
$\omega=v_F\sqrt{2eB/\hbar}$, the lowest band gap $\Delta_{s\xi}\equiv(\Delta_z-s\xi\Delta_{\mathrm{so}})/2$ and  the constants 
$A_{n}^{s\xi}$ and $B_{n}^{s\xi}$ are given by \cite{stille2012}
\begin{eqnarray}
 A_{n}^{s\xi}&=&\left\{\begin{array}{l}
 \mathrm{sgn}(n)\sqrt{\frac{|E_{n}^{s\xi}|+\mathrm{sgn}(n)\Delta_{s\xi}}{{2|E_{n}^{s\xi}|}}},  \quad n\neq 0, \\
\xi_-,  \quad n=0, 
\end{array}\right.\nonumber\\ 
 B_{n}^{s\xi}&=&\left\{\begin{array}{l}
\sqrt{\frac{|E_{n}^{s\xi}|-\mathrm{sgn}(n)\Delta_{s\xi}}{{2|E_{n}^{s\xi}|}}},  \quad n\neq 0, \\
\xi_+,  \quad n=0, 
\end{array}\right.\label{coef}
\end{eqnarray}
The vector  $||n|\rangle$ denotes an orthonormal Fock state of the harmonic
oscillator.

We will do all the numerical analysis in silicene but the results will be  valid just replacing the corresponding parameters in Table \ref{tabla}.

\begin{figure}
\begin{center}
\includegraphics[width=8cm]{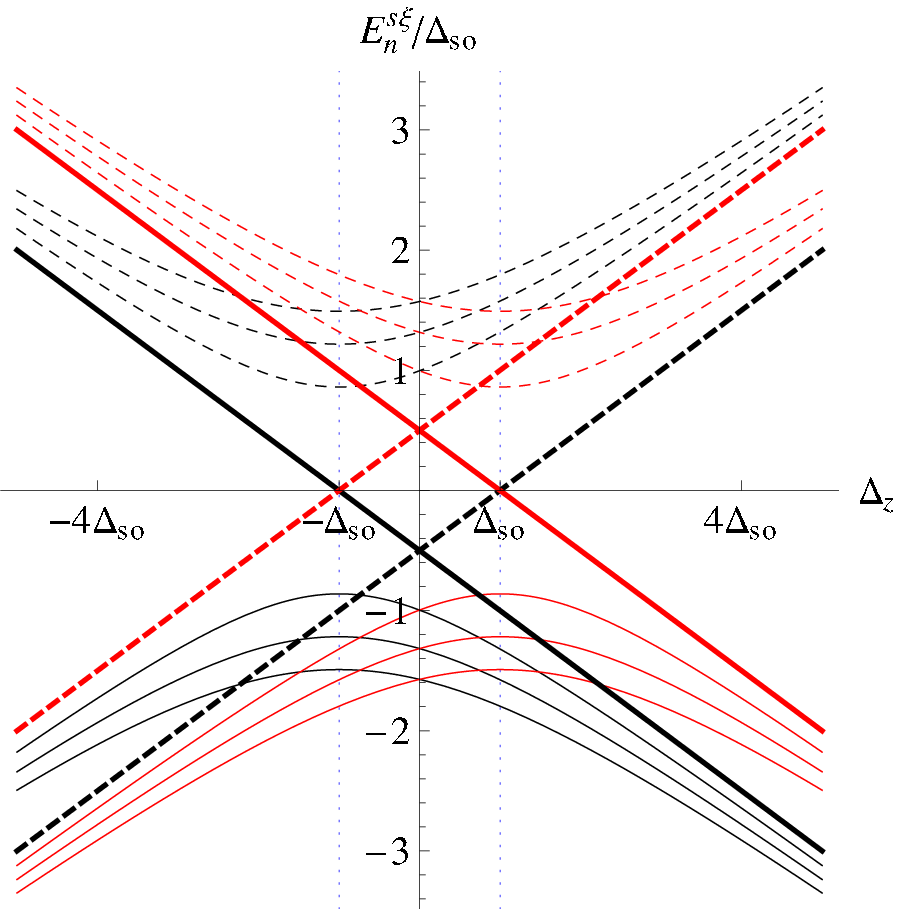}
\end{center}
\caption{Low energy spectra of silicene as a function of the external electric potential $\Delta_z$ for $B=0.05$ T. 
Landau levels $n=\pm 1, \pm 2$ and $\pm 3$, at valley $\xi=1$,  are represented by 
dashed (electrons) and solid (holes) thin lines, black for $s=-1$ and red for
$s=1$ (for the other valley we simply have $E_{n}^{s,-\xi}=E_{n}^{-s,\xi}$). The lowest Landau level $n=0$ is represented by thick lines at both valleys:  solid at $\xi=1$ and dashed 
at $\xi=-1$. Vertical blue dotted grid lines indicate the CNPs separating BI ($|\Delta_z|>\Delta_\mathrm{so}$) from TI   ( $|\Delta_z|<\Delta_\mathrm{so}$) phases.}
\label{energias}
\end{figure}

As already stated, there is a prediction (see e.g. \cite{drummond2012,liu2011,liub2011,Ezawa}) that when the gap $|\Delta_{s\xi}|$ vanishes at the CNP $\Delta_z^{(0)}$, 
silicene undergoes a phase transition from a topological insulator (TI,  $|\Delta_z|<\Delta_\mathrm{so}$) to a band insulator (BI, $|\Delta_z|>\Delta_\mathrm{so}$). 
This topological phase transition entails an energy band inversion. Indeed, in Figure \ref{energias} we show the low energy spectra \ref{especteq} 
as a function of the external electric potential $\Delta_z$ for $B=0.05$ T. One can see that there is a band inversion 
for the $n=0$ Landau level (either for spin up and down) at both valleys. The
energies $E_0^{1,\xi}$ and $E_0^{-1,\xi}$ have the same sign in the BI phase and different sign in the TI phase, thus distinguishing both regimes. 
We will observe a similar ``inversion'' behavior in entropic and variance-based uncertainty relations for the Hamiltonian eigenstates \ref{vectors}, thus providing a quantum-information characterization of the topological 
phase transition.

\section{Entropic uncertainty relation and topological phase transition} \label{entropysec}
\begin{figure}
\begin{center}
\includegraphics[width=8cm]{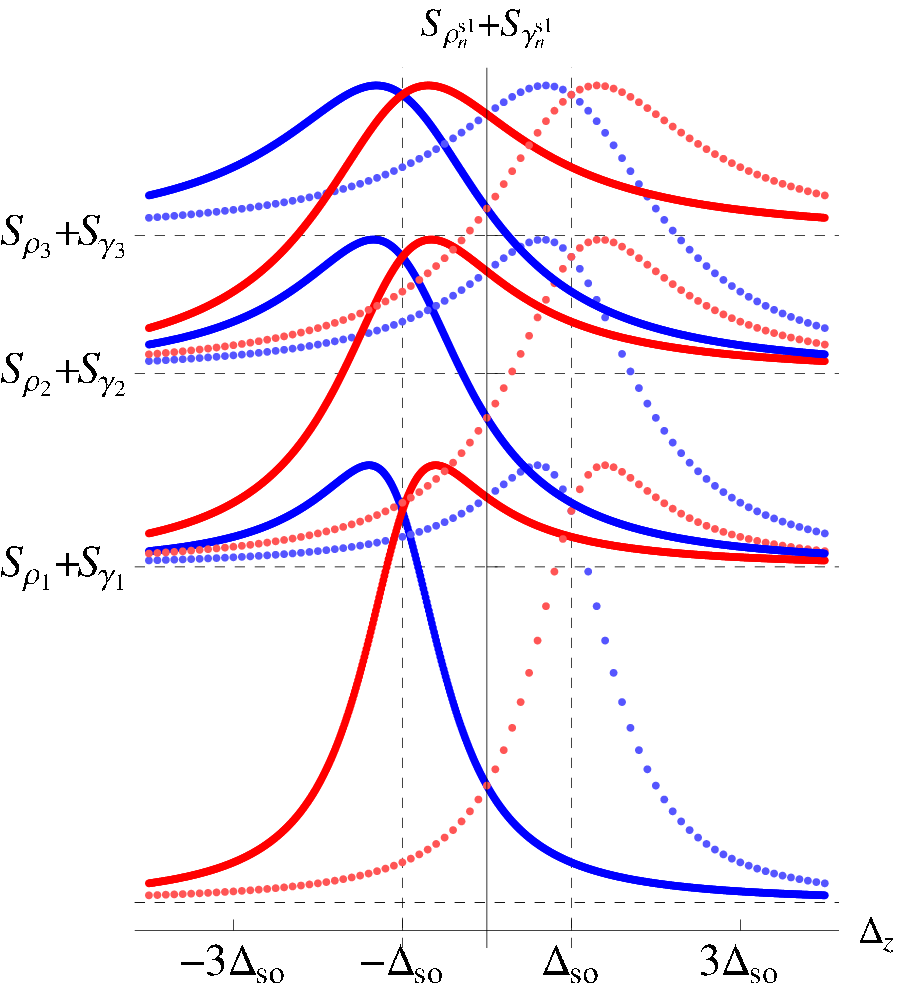}
\end{center}
\caption{ \label{uncer}   Entropic uncertainty relation  $S_{\rho_{n}^{s 1}}+S_{\gamma_{n}^{s 1}}$ 
as a function of the electric potential $\Delta_z$ for the Landau levels: $n=\pm 1, \pm 2$ and $\pm 3$  (electrons in blue  
and holes in red), with spin up (dotted lines) and down (solid lines)
 and  magnetic field $B=0.01$T at valley $\xi=1$. Electron and hole entropy curves 
cross  at the critical value of the electric potential $\Delta_z^{(0)}=-s\Delta_{\mathrm{so}}$ (vertical black dotted grid lines indicate this CNPs), 
$S_{\rho_{|n|}}+S_{\gamma_{|n|}}$ are the asymptotic values for   $\Delta_z\rightarrow\pm\infty$ (horizontal black dotted grid lines).}
\end{figure}
In order to compute Shannon entropies, firstly we have to write the Hamiltonian eigenstates \ref{vectors} in position and momentum representation. 
We know that Fock (number) states $|n\rangle$ can be written in position and momentum representation as
\begin{equation}
 \langle x|n\rangle= \frac{\omega^{1/4}}{\sqrt{2^n n!\sqrt{\pi}}}e^{-\omega x^2/2} H_n\left(\sqrt{\omega} x\right)
\end{equation}
\begin{equation}
\langle p|n\rangle= \frac{(-i)^n}{\sqrt{ 2^n n!\sqrt{\omega\pi}}}e^{- p^2/2\omega} H_n\left(p/\sqrt{\omega} \right)
\end{equation}
where $H_n(x)$ are the Hermite polynomials of degree $n$. We will introduce the number-state densities in position and momentum
 spaces as $\rho_n(x)=|\langle x|n\rangle|^2$ and $\gamma_n(x)=|\langle p|n\rangle|^2$, which are normalized according to $\int \rho_n(x)dx=1$ and $\int \gamma_n(x)dx=1$.
Now, taking into account  Eq. (\ref{vectors}), the position and momentum densities for the Hamiltonian eigenvectors \ref{vectors} are given, respectively, by
\begin{equation}
 \rho_{n}^{s\xi}(x)=(A_{n}^{s\xi})^2|\langle x||n|-\xi_+\rangle_{s\xi}|^2 +(B_{n}^{s\xi})^2|\langle x||n|-\xi_-\rangle_{s\xi}|^2 
\end{equation}
\begin{equation}
\gamma_{n}^{s\xi}(p)=(A_{n}^{s\xi})^2|\langle p||n|-\xi_+\rangle_{s\xi}|^2 +(B_{n}^{s\xi})^2|\langle p||n|-\xi_-\rangle_{s\xi}|^2.
\end{equation}
We will study the position and momentum entropies 
\begin{equation}
 S_{\rho_{n}^{s\xi}}\equiv -\int_{-\infty}^{\infty} \rho_{n}^{s\xi}(x)\ln\left(\rho_{n}^{s\xi}(x)\right) dx
\end{equation}
\begin{equation}
  S_{\gamma_{n}^{s\xi}}\equiv -\int_{-\infty}^{\infty} \gamma_{n}^{s\xi}(p)\ln\left(\gamma_{n}^{s\xi}(p)\right) dp.
\end{equation}
If we make a change of variable, it is straightforward to see that  $S_{\rho_{n}^{s\xi}}=S_{\gamma_{n}^{s\xi}}-\ln\left(\omega\right)$.

\begin{figure}
\begin{center}
\includegraphics[width=6cm]{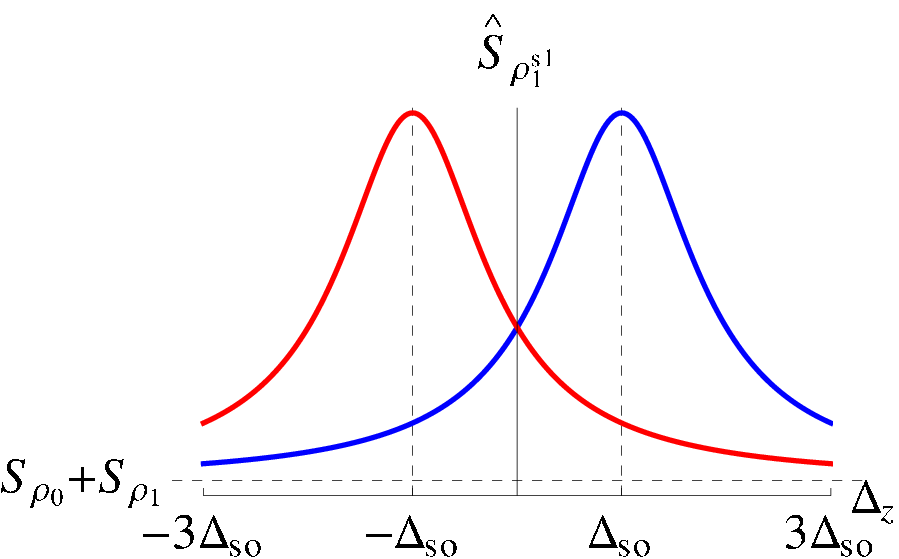}
\includegraphics[width=6cm]{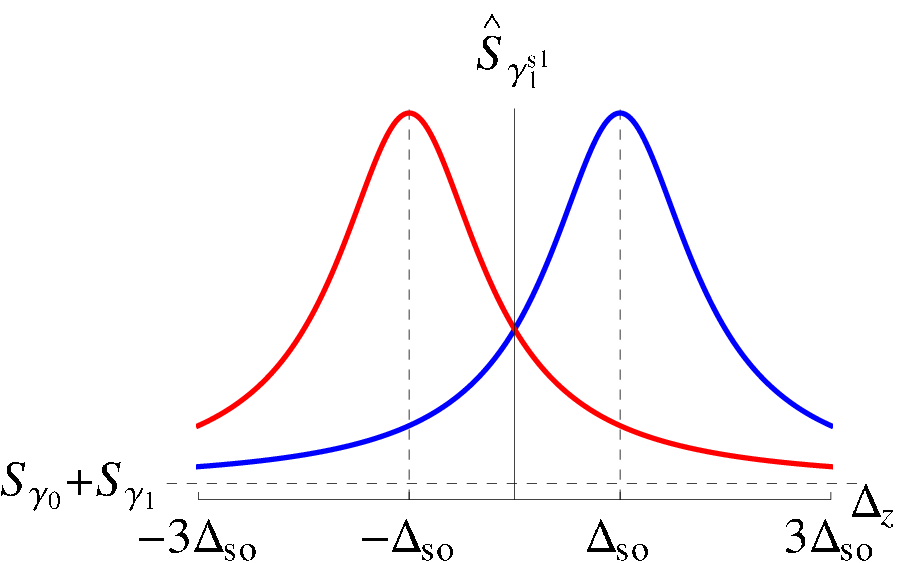}
\end{center}
\caption{ \label{sumehpomo} Combined Shannon entropies \ref{comb1} and \ref{comb2} of electron plus holes for $n=1$, in position and momentum spaces, $\hat{S}_{\rho_{n}^{s 1}}$ (left) and $\hat{S}_{\gamma_{n}^{s 1}}$ (right), respectively, 
as a function of the electric potential $\Delta_z$, for $B=0.01$T,  with spin up (blue) and down (red)
 and valley $\xi=1$. The combined entropies have a maximum 
  at the critical value of the electric potential $\Delta_z^{(0)}=-s\Delta_{\mathrm{so}}$ (vertical dashed grid lines
 indicate these CNPs).}
\end{figure}

In figure  \ref{uncer} we plot $S_{\rho_{n}^{s\xi}}+ S_{\gamma_{n}^{s\xi}}$ as a function of the external electric potential
$\Delta_z$ for the Landau levels: $n=\pm 1, \pm 2$ and $\pm 3$  (electrons in blue  
and holes in red), with spin up (dotted lines) and down (solid lines)
 and  magnetic field $B=0.01$T at valley $\xi=1$.
 Electron-hole entropy curves cross at the CNP
  $|\Delta_z|=\Delta_\mathrm{so}$.
For electrons (resp. holes), the asymptotic entropies in position space  are given by
 $S_{\rho_{|n|-1}}$ (resp. $S_{\rho_{|n|}}$)  for $\Delta_z\rightarrow\infty$ and $S_{\rho_{|n|}}$ (resp. $S_{\rho_{|n|-1}}$)  for $\Delta_z\rightarrow-\infty$. 
In momentum space the behavior is analogous.
The uncertainty relation has the same value for spin up (resp. down) electrons and holes
at the CNP point $\Delta_z=\Delta_\mathrm{so}$ (resp. $\Delta_z=-\Delta_\mathrm{so}$).
Moreover, for each $n$, note that  in the BI phase  the electrons (resp. holes) uncertainty goes to  greater value than holes (resp. electrons) uncertainty
 for $\Delta_z<-\Delta_\mathrm{so}$ 
(resp. $\Delta_z>\Delta_\mathrm{so}$). We have checked that the smaller the magnetic field strength, the sharper this effect is.

In figure \ref{sumehpomo} we have plotted the combined entropy of electrons plus holes
\begin{equation}
 \hat{S}_{\rho_{n}^{s 1}}=S_{\rho_{n}^{s 1}}+ S_{\rho_{-n}^{s 1}}\label{comb1}
\end{equation}
\begin{equation}
 \hat{S}_{\gamma_{n}^{s 1}}=S_{\gamma_{n}^{s 1}}+ S_{\gamma_{-n}^{s 1}}\label{comb2}
\end{equation}
in position and momentum representation for $n=1$. We can observe that the combined entropies exhibit a maximum at the CNPs in both representation spaces. 
This is a common feature for general $n$.

\section{Kullback-Leibler entropy}\label{KLsection}
\begin{figure}
\begin{center}
\includegraphics[width=8cm]{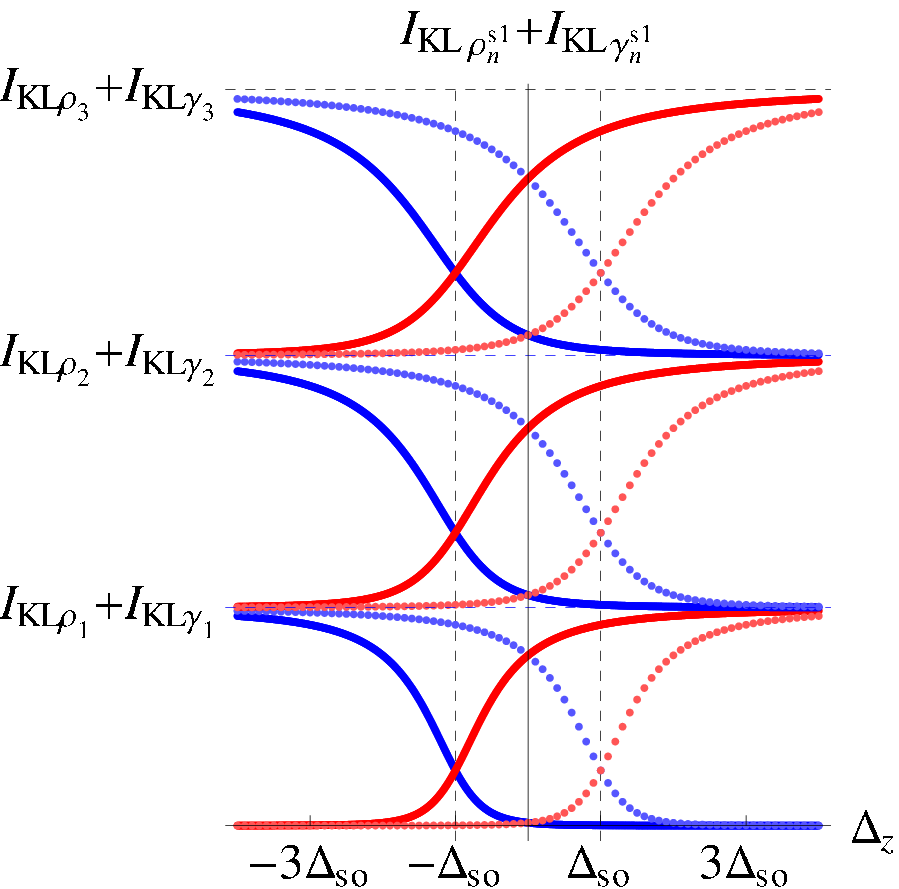}
\end{center}
\caption{ \label{KLuncer} Kullback-Leibler  uncertainty  relation  $I_{KL\rho_{n}^{s 1}}+I_{KL\gamma_{n}^{s 1}}$ 
as a function of the electric potential $\Delta_z$ for the Landau levels: $n=\pm 1, \pm 2$ and $\pm 3$  (electrons in blue  
and holes in red), with spin up (dotted lines) and down (solid lines)
 and  magnetic field $B=0.01$T at valley $\xi=1$. Electron and hole entropy curves 
cross  at the critical value of the electric potential $\Delta_z^{(0)}=-s\Delta_{\mathrm{so}}$ (vertical black dotted grid lines indicate this CNPs), 
}
\end{figure}
The relative  or Kullback-Leibler entropy is a measure for the deviation of a density $f({\bf r})$ from a reference density $g({\bf r})$ \cite{KL} is defined as
\begin{equation}
I_{KL}(f,g)=\int f({\bf r})\ln \left(\frac{f({\bf r})}{g({\bf r})}\right)d{\bf r}
\label{Kullback}
\end{equation}
Recently, the relative R\'enyi and Kullback-Leibler entropies have been found to be an excellent marker of a quantum phase transition in the Dicke \cite{JSM,Nagy} and 
in the $U(3)$ vibron model \cite{JMM}. In this section we will explore the utility of the relative entropy as an indicator of a topological phase transition. 
For this purpose, we will consider as  reference densities the  position and momentum densities $\rho_0(x)$ and $\gamma_0(p)$, respectively,
 which are the densities for minimum uncertainty in relation (\ref{entropic}). Therefore, we will analyze how different is a density from the minimum uncertainty density.

We will study the position and momentum entropies 
\begin{equation}
 I_{KL\rho_{n}^{s\xi}}\equiv \int_{-\infty}^{\infty} \rho_{n}^{s\xi}(x)\ln\left(\frac{\rho_{n}^{s\xi}(x)}{\rho_0(x)}\right) dx
\end{equation}
\begin{equation}
  I_{KL\gamma_{n}^{s\xi}}\equiv \int_{-\infty}^{\infty} 
\gamma_{n}^{s\xi}(p)\ln\left(\frac{\gamma_{n}^{s\xi}(p)}{\gamma_0(p)} \right)dp. 
\end{equation}
Again, it is straightforward that  $I_{KL\rho_{n}^{s\xi}}=I_{KL\gamma_{n}^{s\xi}}$.


In figure   \ref{KLuncer} we plot  the sum $I_{KL\rho_{n}^{s 1}}+ I_{KL\gamma_{n}^{s 1}}$ as a function of the external electric potential
$\Delta_z$ for the Landau levels $n=\pm 1, \pm 2$ and $\pm 3$  (electrons in blue  
and holes in red), with spin up (dots lines) and down (solid lines).
 The figure corresponds to a magnetic field $B=0.01$T. Electron-hole relative entropy curves cross at the CNP
  $|\Delta_z|=\Delta_\mathrm{so}$ at which they reach the values  $\bar{I}_{n}^s\approx 0.37$, $1.96$ and $3.69$ for $|n|=$ $1$, $2$ and $3$ respectively.
For electrons (resp. holes) the asymptotic entropies in position space  are given by
 $I_{KL\rho_{|n|-1}}$ (resp. $I_{KL\rho_{|n|}}$)  for $\Delta_z\rightarrow\infty$ and $I_{KL\rho_{|n|}}$ (resp. $I_{KL\rho_{|n|-1}}$)  for $\Delta_z\rightarrow-\infty$. 
In momentum space the behavior is analogous. Note the electron-hole entropy inversion phenomenon anticipated at the end of Section \ref{Hamilsec}. Indeed, the quantity $I_{KL\rho_{n}^{s 1}}+ I_{KL\gamma_{n}^{s 1}}-\bar{I}_{n}^s$ 
has the same sign for spin up and down electrons 
(idem for holes) in the BI phase ($|\Delta_z|>\Delta_\mathrm{so}$) and different sign in the TI phase ($|\Delta_z|<\Delta_\mathrm{so}$).

\section{Heisenberg uncertainty relation and topological phase transition}\label{numsec}
\begin{figure}
\begin{center}
\includegraphics[width=8cm]{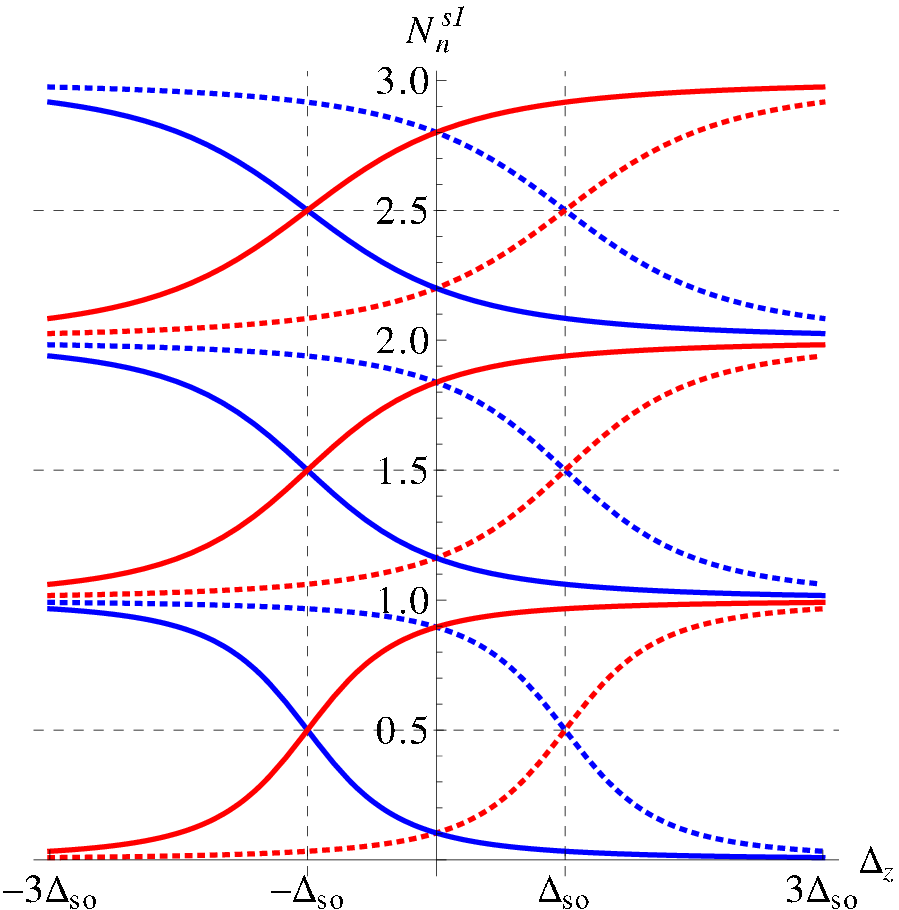}
\end{center}
\caption{ \label{numfig}   Expectation value of the number operator $N=a^\dag a$ in the energy eigenstate $|n\rangle_{s\xi}$ 
as a function of the electric potential $\Delta_z$ for the Landau levels: $n=\pm 1, \pm 2$ and $\pm 3$  (electrons in blue  
and holes in red), with spin up (dotted lines) and down (solid lines)
 and  magnetic field $B=0.01$T at valley $\xi=1$. Mean number curves 
cross  at the critical value of the electric potential $\Delta_z^{(0)}=-s\Delta_{\mathrm{so}}$ 
(vertical black dotted grid lines indicate these CNPs), for which $N_n^{s\xi}=|n|/2$ (horizontal black dotted grid lines)}
\end{figure}
Entropic uncertainty relation provides  a
refined version of the Heisenberg uncertainty relation (see \cite{Halliwell} and references therein):
\begin{eqnarray}
\label{bbrud}
\Delta x \Delta p  \ge \frac12 \exp{[S_{\rho} + S_{\gamma} -1 - \ln{\pi}]} \ge \frac12 .
\end{eqnarray}
It gives a stronger bound for the  variance product than the standard $\frac12$. In this section, we shall explore the more usual 
Heisenberg uncertainty relation.

Introducing position and momentum operators through a bosonic mode, $[a,a^\dag]=1$, as usual:
\begin{equation} X=\frac{1}{\sqrt{2\omega}}(a^\dag+a), \; P=i\sqrt{\frac{\omega}{2}}(
a^\dag-a),\label{posmomenop}
\end{equation}
we can easily compute the expectation values of $X$ and $P$ and their fluctuations in an energy eigenstate \ref{vectors} as
\begin{eqnarray}
&\langle n|X|n\rangle_{s\xi}=0,\; \langle n|P|n\rangle_{s\xi}=0,&\\
&\langle n|X^2|n\rangle_{s\xi}=\frac{1}{\omega}(N_n^{s\xi}+\frac12)=\frac{1}{\omega^2}
\langle n|P^2|n\rangle_{s\xi},&
\end{eqnarray}
where 
\begin{equation}
 N_n^{s\xi}=(A_n^{s\xi})^2(|n|-1)+(B_n^{s\xi})^2|n|\label{numeq}
\end{equation}
is the expectation value of the number operator $N=a^\dag a$ in the energy eigenstate $|n\rangle_{s\xi}$. Therefore, the 
product of standard deviations, $\Delta X_n^{s\xi} \Delta P_n^{s\xi}=N_n^{s\xi}+\frac12$, is written in terms of 
$N_n^{s\xi}$ solely. In Figure \ref{numfig} we represent 
$N_n^{s\xi}$ as a function of the electric potential $\Delta_z$ for several Landau levels. We observe the same electron-hole 
curve inversion phenomenon as for the entropy curves. The quantity $N_n^{s\xi}-\frac{|n|}{2}$ has the same sign for spin up and down electrons 
(idem for holes) in the BI phase ($|\Delta_z|>\Delta_\mathrm{so}$) and different sign in the TI phase ($|\Delta_z|<\Delta_\mathrm{so}$).

\section{Conclusions}
We have explored how different quantifications  of the uncertainty relation characterize a topological phase
 transition in  a group of 
2D Dirac gapped materials (monolayer sheets of Si, Ge, Sn, and Pb). Firstly we have  inspected the entropic uncertainty relation.
 We have found that the electron-hole entropic
 curves cross at the charge neutrality poin (CNP), that is, the electron and holes with spin up (down) have  the same uncertainty at the critical point 
$\Delta_z=\Delta_\mathrm{so}$  ($\Delta_z=-\Delta_\mathrm{so}$). The combined
entropy of electrons plus holes shows  a maximum at the critical points in position and momentum spaces; therefore, 
the combined electron plus hole density, for spin up (resp. down), is more delocalized at  the CNP $\Delta_z=\Delta_\mathrm{so}$  
(resp. $\Delta_z=-\Delta_\mathrm{so}$).
 Furthermore, we have analyzed the relative (or Kullback-Leibler) entropy,  finding that the sum of the relative entropies in
position and momentum spaces identifies each phase (band and topological insulator) at the CNPs.
For completeness we have considered the product of standard deviations observing the same electron-hole uncertainty curve inversion phenomenon.
Summarizing, we have related the uncertainty principle and the topological phase transitions in this model. 
We expect  that this analysis might be applicable to other problems to obtain a general/deepest connection between both 
concepts: uncertainty principle and topological phase transitions.

\section*{Acknowledgments}
  The work was supported by 
the Spanish Projects: MICINN
FIS2011-24149, CEIBIOTIC-UGR PV8 and the Junta de Andaluc\'{\i}a projects FQM.1861 and FQM-381.


\begin{thebibliography}{100}



\bibitem{takeda}
K. Takeda, K. Shiraishi,
Phys. Rev. B 50 (1994) 14916.

\bibitem{guzman94}
G. G. Guzman-Verri, L. Lew Yan,
Phys. Rev. B 76 (2007) 075131.

\bibitem{vogt12}
P. Vogt et al., Phys. Rev. Lett.  108 (2012) 155501.

\bibitem{Aufray10}
B. Augray, A. Kara, S. B. Vizzini, H. Oughaldou,
C. L\'eAndri, B. Ealet, G. Le Lay,
App. Phys. Lett. 96 (2010) 183102.

\bibitem{lalmi10}
B. Lalmi, H. Oughaddou, H. Enriquez, A. Kara, S. B.   Vizzini, B. N.
Ealet, B. Augray,
App. Phys. Letters 97 (2010) 223109.

\bibitem{fleurence12}
A. Feurence, R. Friedlein, T. Ozaki, H. Kawai, Y. Wang, Y. Y. Takamura,
Phys. Rev. Lett. 108 (2012) 245501.

\bibitem{padova}
P. E. Padova et al.,
App. Phys. Lett. 96 (2010) 261905. 

\bibitem{nature} W-F. Tsai, C-Y. Huang, T-R Chang et al. Nat. Commun. {\bf 4}, 1500 (2013) 

\bibitem{tahir2013} M. Tahir, U. Schwingenschl\"ogl, Scientific Reports, {\bf
  3}, 1075 (2013).

\bibitem{Kane} Kane C L and Mele E J,  Phys. Rev. Lett. {\bf 95}, 226801 (2005)

\bibitem{Bernevig} B. Andrei Bernevig, Taylor L. Hughes and Shou-Cheng Zhang, Science {\bf 314}, 1757-1761 (2006).  

\bibitem{calixto14} M. Calixto and E. Romera, EPL {\bf 109},  40003 (2015)


\bibitem{10} I. I. Hirschman, Am. J. Math. {\bf 79}, 152 (1957)

\bibitem{11} Bialynicki-Birula, I., Mycielski, J.: Commun. Math. Phys. {\bf 44}, 129 (1975)
\bibitem{21}W. Beckner, Ann. Math. {\bf 102}, 159 (1975)

\bibitem{15} H. Maassen, J. B. M.  Uffink, Phys. Rev. Lett. {\bf 60}, 1103 (1988)
\bibitem{16} J. S\'anchez-Ruiz,  Phys. Lett. A {\bf 244}, 189 (1998)
\bibitem{Majernik} E. Majern\'\i kov\'a, V. Majern\'\i k, and S. Shpyrko, Eur. Phys. J. B
{\bf 38}, 25 (2004).


\bibitem{17} I. Bialynicki-Birula, Phys. Rev. A {\bf 74}, 052101 (2006)
\bibitem{18} E. Romera, F. de los Santos Phys. Rev. Lett.{\bf  99}, 263601 (2007); Phys Rev. A 78, 013837 (2008).
\bibitem{found} T. Sch\"urmann and I. Hoffmann Found Phys {\bf 39}, 958 (2009).
\bibitem{gadre} S. R. Gadre, Phys. Rev. A {\bf 30}, 620 (1984).
\bibitem{sears} S. R. Gadre, S. B. Sears, S. J. Chakrovarty and R. D. Bendale
  Phys. Rev. A {\bf 32}, 2602 (1985).
\bibitem{romera12} E. Romera, M. Calixto and \'A. Nagy, EPL, {\bf 97}, 20011 (2012).
\bibitem{calixto12} M. Calixto, \'A. Nagy, I. Paradela and E. Romera, Phys. Rev. A {\bf 85}, 053813 (2012)

\bibitem{drummond2012} N. D. Drummond, V. Z\'olyomi, and V. I. Fal'ko, Phys. Rev. B {\bf 85}, 075423 (2012).
\bibitem{liu2011} C. C. Liu, W. Feng, and Y. Yao, Phys. Rev. Lett. {\bf 107}, 076802 (2011).
\bibitem{liub2011} C. C. Liu, H. Jiang, and Y. Yao, Phys. Rev. B {\bf 84}, 195430 (2011).
\bibitem{fermispeed1} S. Trivedi, A. Srivastava and R. Kurchania, J. Comput. Theor. Nanosci. {\bf 11}, 1-8 (2014). doi:10.1166/jctn.2014.3428
\bibitem{fermispeed2} B van den Broek et al. 2D Materials {\bf 1}
(2014) 021004. doi:10.1088/2053-1583/1/2/021004
\bibitem{stille2012} L. Stille, C. J. Tabert, and E. J. Nicol, Phys. Rev. B {\bf 86}, 195405 (2012); 
C.J. Tabert and E.J. Nicol, Phys. Rev. Lett. {\bf 110}, 197402 (2013); C.J. Tabert and E.J. Nicol, Phys. Rev. B {\bf 88},
085434 (2013).
 
 \bibitem{Ezawa} M. Ezawa, New Journal of Physics {\bf 14}
(2012) 033003




\bibitem{KL} S. Kullback and R. A. Leibler, Ann. Math. Stat. {\bf 22}, 79 (1951).

\bibitem{JSM} E. Romera, K. Sen, \'A. Nagy, J. Stat. Mech. P09016 (2011).
\bibitem{Nagy} E. Romera and \'A. Nagy, Phys. Lett. A {\bf 377}, 3098 (2013). 
\bibitem{JMM} E. Romera, M. Calixto and \'A. Nagy, J.  Mol. Model. {\bf 20}, 2237 (2014)

\bibitem{Halliwell} J. J. Halliwell, Phys. Rev. D {\bf 48}, 2739 (1993).




\end{thebibliography}
\end{document}